# Beam Based RF Voltage Measurements & Longitudinal Beam Tomography at the Fermilab Booster

C. M. Bhat and S. Bhat
*Fermi National Accelerator Laboratory, Batavia IL 60510, USA*

Increasing proton beam power on neutrino production targets is one of the major goals of the Fermilab long term accelerator programs. In this effort, the Fermilab 8 GeV Booster synchrotron plays a critical role for at least the next two decades. Therefore, understanding the Booster in great detail is important as we continue to improve its performance. For example, it is important to know accurately the available RF power in the Booster by carrying out beam-based measurements in order to specify the needed upgrades to the Booster RF system. Since the Booster magnetic field is changing continuously measuring/calibrating the RF voltage is not a trivial task. Here, we present a beam based method for the RF voltage measurements. Data analysis is carried out using computer programs developed in Python and MATLAB. The method presented here is applicable to any RCS which do not have *flat-bottom* and *flat-top* in the acceleration magnetic ramps. We have also carried out longitudinal beam tomography at injection and extraction energies with the data used for RF voltage measurements. Beam based RF voltage measurements and beam tomography were never done before for the Fermilab Booster. The results from these investigations will be very useful in future intensity upgrades.

## 1. INTRODUCTION

For the past one and a half decades Fermilab has been making significant upgrades to all accelerators in the complex to meet the high intensity proton demands by the intensity frontier experiments – short and long-baseline neutrino experiments and experiments with muon beams. The first part of these upgrades was "Proton Improvement Plan-I" [1] (referred to as PIP, now denoted PIP-I). The baseline goals of PIP-I were to provide ~4.3E12 protons per Booster cycle (ppBc) at 8 GeV from the Booster synchrotron at 15 Hz, deliver 700 kW proton beam power on the NuMI/NOvA neutrino target, and at the same time send ~4.6E16 p/hr to the Booster neutrino experiments. The Booster is a 400 MeV- 8 GeV proton synchrotron, one of the earliest built rapid cycling synchrotrons (RCS) in the world. Figure 1 displays an aerial view of the Fermilab Booster and its 400 MeV H$^-$ Injector LINAC. Until recently beam was accelerated only on twelve of its fifteen cycles. Even though PIP-I is scheduled to be completed by 2018, many of its goals have been met already, including 700 kW proton beam power on the NuMI/NOvA neutrino target, nearly a year ahead of the scheduled date of completion. The Booster is also at full 15 MHz beam operation. The next major upgrade to the accelerator complex called PIP-II [2], planned to be completed by 2028, has the following primary goals 1) replace the existing normal conducting RF LINAC by a superconducting RF LINAC, 2) increase the Booster cycle rate from 15 Hz to 20 Hz and 3) achieve beam power >1.2 MW on the LBNF neutrino target. The Booster, therefore, will continue to play a very significant role in the Fermilab accelerator based HEP programs well beyond PIP-II.

During 2014-15, a novel technique for beam injection and capture into the Booster, called "early beam injection scheme" [3], was introduced. (This was outside the scope of PIP-I.) This improved the quality of the beam considerably and beam intensity to the downstream neutrino experiments beyond the PIP-I goals. A schematic of the early injection scheme is shown in Fig. 2. The longitudinal beam dynamics simulations for the scheme showed that one can achieve 40% higher beam intensity than in the past with good quality beam with no beam loss. Further, near-term improvements in the Booster, Recycler, Main Injector and neutrino targets can provide substantial increase in beam power on the neutrino target while preparing the accelerator complex for high intensity operation with PIP-II. Therefore, an intermediate initiative called PIP-I+ has been undertaken. A comparison between various parameters of PIP-I, PIP-I+ and PIP-II projects is given in Table -I.

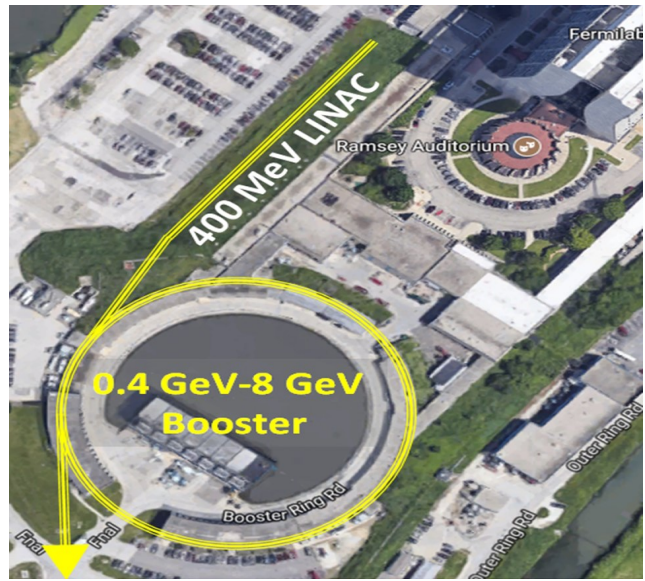

Figure 1: Aerial view of the Fermilab Booster complex and 400 MeV LINAC.

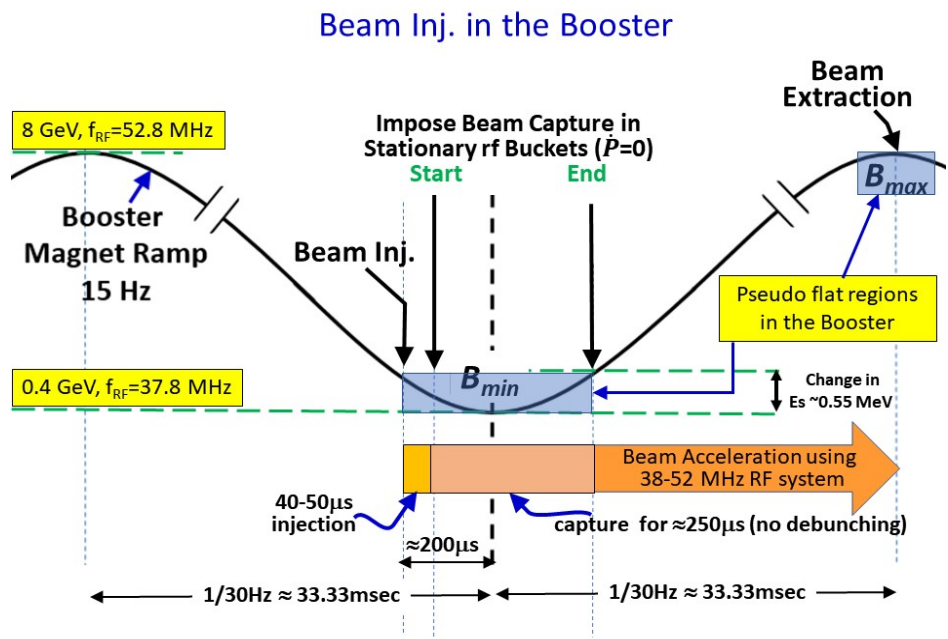

Figure 2: Schematic view of the early injection scheme with sinusoidal magnetic ramp of dipole magnets in the Booster. Beam injection and extraction regions are also indicated. Pseudo flat regions are created at both ends.

Accelerators/ Beam Based RF Voltage Measurements & Longitudinal Beam Tomography …

Table I: PIP-I, PIP-I+ and PIP-II parameters.

| Parameters | PIP-I (2018) | PIP-I+ ⊗ | PIP-II (2028) |
|---|---|---|---|
| Inj. & Extraction K.E. (GeV) | 0.4, 8.0 | 0.4, 8.0 | 0.8, 8.0 |
| Injector to the Fermilab Booster | Current LINAC | Current LINAC | New SC LINAC |
| Inj. & Extraction ppBc (xE12) | 4.62, 4.3 | 5.2, 4.93 | 6.7, 6.5 |
| Number of Booster Turns | 12-14 | up to 22 | 300 |
| Efficiency (%) | 92 | ≈95♣ | 97 |
| Booster repetition rate (Hz) | 15 | 15 | 20 |
| Beam Power at Extraction (kW) | 82 | 95 | 166 |
| Booster batches for MI | 12/1.33 sec | 12/1.2 sec | 12/1.2 sec |
| RR/MI Beam efficiency (%) | 95 | 95 | 97 |
| NOvA beam power (kW) | 700 | ~900 | 1200 ⁋ |
| Booster Beam for other users (kW) | 33 | 32 | 83 |
| Booster flux (protons/hr) | ~ 2.3E17 | ~2.7E17 | ~ 4.6E17 |
| Booster Vrf (MV) | 1 | 1.1 | 1.3 |
| Number of Booster RF cavities | 20 | 22 | -- |
| Beam loading/Booster cavity (kV) | 12 | 16 | 18 |

⊗ Booster beam intensity at extraction >5.5E12 ppBc has been demonstrated. Assuming upgrades in Booster, RR & MI required by PIP-II are done ahead of time, except that the PIP-I+ uses current 400 MeV LINAC with longer LINAC beam pulses.

♣ Assuming the same number of proton loss/Booster cycle as expected during PIP.

⁋ Upgrade potential for PIP-II is 2400 kW.

With increase in beam intensity, beam loading in the RF system becomes an issue and the required RF power will also increase. The Booster RF cavities operate in the frequency range of 38-52 MHz and are nearly 45 years old. As a part of the PIP-I upgrades, its twenty RF cavities were refurbished and two new cavities were added. Each cavity can provide up to 50 kV of RF voltage. Figure 3 shows a typical Booster RF cavity with its power amplifier and distribution of all twenty-two RF cavities in the ring. These cavities are configured in two groups "A" and "B", so that two consecutive cavities belong to different groups. This arrangement is necessary for a small size ring like the Booster and enables us to treat each group as vectors and apply net RF voltage in the range from zero (phase angle between vectors=$180^0$) to a maximum (phase angle =$0^0$) value without turning off any RF amplifiers. PIP-I and PIP-I+ demand a total RF voltage of 1 MV and 1.1 MV, respectively for beam operation (for PIP-II the RF power demand is higher and some of the RF cavities must be replaced with higher gradient cavities). But, it is quite critical to ensure that their placement in the ring and phase between successive cavities provide the required RF voltage. Since the RF frequency changes by ~ 40% from injection to extraction energy, adjusting the phase between the RF cavities is not easy. Thus, a beam based RF voltage calibration becomes essential to fulfill RF requirements for future upgrades.

Beam based RF calibration is not a trivial task in the case of Booster. In this paper, we present two different techniques for such a calibration. Both use the same principle.

Accelerators/ Beam Based RF Voltage Measurements & Longitudinal Beam Tomography …

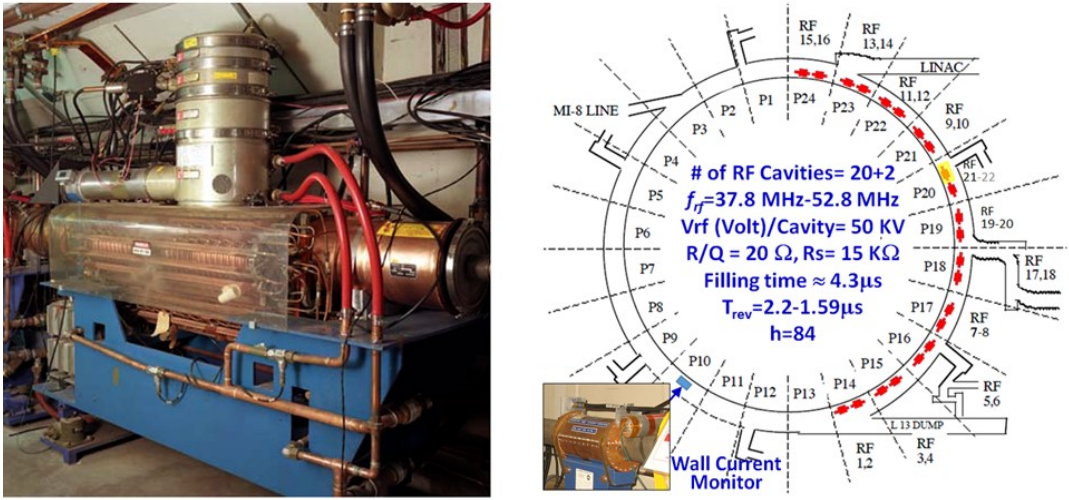

Figure 3: An example of a refurbished Booster RF cavity (left) and schematic view of the Fermilab Booster ring indicating the distributions of RF cavities (right). Location of the wall current wall monitor used for measuring the line-charge distributions of the beam pulses in the Booster is also shown.

## 2. PRINCIPLE OF BEAM BASED RF VOLTAGE MEASUREMENTS

The synchrotron frequency of a beam particle, $f_{sy}(\hat{\phi})$, in a stationary bucket produced by RF voltage $Vrf$, is given by,

$$f_{sy}(\hat{\phi}) = \pi f_{rev} \sqrt{\frac{heVrf|\eta|}{2\pi\beta^2 E_s}} \bigg/ 2K(\sin(\hat{\phi}/2)) \qquad (1)$$

where $\hat{\phi}$ is the maximum amplitude of synchrotron oscillation as shown in Fig 4(a). The quantity $K$ is complete elliptical integral of the first kind, given by,

$$K(x) = \int_0^{\frac{\pi}{2}} \frac{dr}{\sqrt{1 - x^2 \sin^2(r)}} \qquad (2)$$

where, $x = \sin(\hat{\phi}/2)$. The quantities $f_{rev}, h, E_s, \eta, e$ and $\beta$ are beam revolution frequency, RF harmonic number, synchronous energy of the beam particle, slip factor at $E_s$, electron charge and relativistic velocity, respectively. In the small angle approximation for $\hat{\phi}$, $K(x) = \pi/2$. Then $Vrf$ in volts can be written as,

$$Vrf = \left(\frac{f_{sy}(\hat{\phi})}{f_{rev}}\right)^2 \frac{2\pi\beta^2 E_s}{eh|\eta|} \qquad (3).$$

Accelerators/ Beam Based RF Voltage Measurements & Longitudinal Beam Tomography …

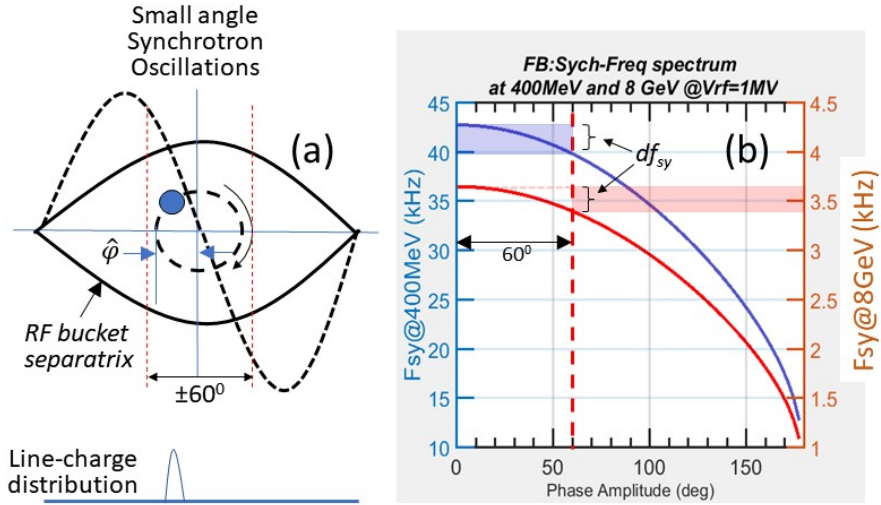

Figure 4: (a) Schematic view of small angle synchrotron oscillations for a beam particle (closed dashed curve), bucket boundary (solid closed curve) and rf wave-form (dashed line). (b) The calculated single particle synchrotron frequency spectrum in a stationary rf bucket in the Booster at 400 MeV (blue curve, left ordinate) and at 8 GeV (red curve, right ordinate). The maximum error in $f_{sy}$ is <7% if $\hat{\phi}<60^0$.

In case of the Fermilab Booster, we have $f_{rev}=451.16\,\text{kHz}, \eta=-0.458, \beta=0.713, E_s=1.338\,\text{GeV}/c^2$ at injection and $f_{rev}=629.22\,\text{kHz}, \eta=0.022, \beta=0.994, E_s=8.938\,\text{GeV}/c^2$, at extraction.

Figure 4 shows a schematic view of the small angle synchrotron oscillations in a stationary RF bucket and calculated synchrotron frequency distributions at 400 MeV and at 8 GeV in the Booster for 1 MV of RF voltage. By measuring small angle synchrotron frequency of the beam, one can get $Vrf$. This requires measurements on line-charge distribution of beam particles in an RF bucket (as shown in Fig. 4(a)) and its time evolution using a wall current monitor (WCM) for a few synchrotron oscillations. Fourier transformation of the measured WCM data gives information on the synchrotron oscillation side band around RF frequency. At synchrotron oscillation amplitude of $\hat{\phi}=60^0$ the maximum uncertainty in $f_{sy}$ is about 7%. In principle, one can measure synchrotron oscillation period of beam particles in an accelerating RF bucket if the phase angle of the beam centroid relative to the RF waveform is also known. Generally, it is very difficult to measure the phase angle with enough accuracy while beam is being accelerated. However, measurements made here on stationary RF buckets give very robust $Vrf$ calibration.

The beam based RF voltage calibration outlined above is a widely known technique for synchrotron storage rings and synchrotron accelerators with a long flat-bottom and/or a flat-top magnetic field regions. In both cases, measurements are made using the beam in stationary buckets. But, in an RCS like the Fermilab Booster the magnetic field is changing continuously and therefore, such measurements are not straight-forward. In the next section, we present our methods for RF voltage measurements.

Accelerators/ Beam Based RF Voltage Measurements & Longitudinal Beam Tomography …

## 3. $f_{sy}$ MEASUREMENTS IN THE BOOSTER

In the Fermilab Booster, we have carried out $f_{sy}$ measurements under two different scenarios. The first measurement uses beam in stationary RF buckets when Booster magnetic field is held at 400 MeV. The second one is carried out on a *pseudo* flat top in changing magnetic field.

### 3.1. 400 MeV DC Mode Measurements

The 400 MeV measurements were made on a DC ramp of the magnetic field of the Booster, i.e., setting $B_{max}-B_{min}=0$ (see Fig. 2). About 100 μs prior to the beam injection the voltages of four RF stations were set at their maximum value with their vector sum set to zero and RF frequency set to the injection frequency. Then about 1E12 protons were injected into the Booster and beam was captured non-adiabatically in about 250 μs by increasing the RF voltage, i.e., by changing the relative phase angle between A and B RF vectors. The sum RF voltage, *RFSUM*, was held at its maximum value for about 20 msec. For the last 14 msec of the beam cycle the RF voltage on each RF station was reduced as shown by the blue curve (which is an input to the Booster low level RF (LLRF)) in Fig. 5. A drop in the *RFSUM* at about 18 msec is because the Booster LLRF forces a jump in the RF phase (the LLRF used here was same as that for beam acceleration that triggers the phase jump from $\hat{\phi}$ to $\pi - \hat{\phi}$ at transition energy around 18 msec from the injection in the cycle). Even though the beam was held at injection energy for the entire beam cycle and no phase jump, the time-based phase jump in the LLRF influences the *RFSUM*. The *RFSUM* is a vector sum of individual RF cavity gap monitor-outputs corrected for beam-loading (at the level of ~20%). The beam intensity measured by a toroid in the 400 MeV beam line between the LINAC and the Booster is shown by dark green curves. The purple curves in Fig. 5 are for beam intensity in the Booster for multiple cycles. The behavior of the beam in the Booster is representative of its lifetime in the ring at low B-field and small *RFSUM*. As the *RFSUM* is increased the lifetime found to improve. The line-charge distribution of the circulating beam is measured using a WCM shown in Fig.3(b). For the RF voltage calibration, we focus only on first 18 msec in the beam cycle. Two sets of WCM data that spans 200 μs each were taken, one at 5 msec and another one either at 10 msec or 15 msec after the beam capture. Beam is extracted at the end of 33.33 msec. These measurements were repeated by turning on two RF stations at a time as shown in Fig. 5.

A Python program was written to analyze the data. FFT of the raw WCM data gives the RF frequency and the revolution period of 2.206 μs. Mountain range and density distributions as a function of beam turn number are shown in Fig. 6. If the beam particle distribution is nearly uniform as shown schematically in Fig. 6(a), then the mountain range and its contour (density) plots appear as illustrated in Fig. 6(d) for last two bunches. In the case of a bunch with non-uniform distribution with a distinct clump, its density plot would look like the first bunch in Fig 6(d). In this case, we can visually observe the sinusoidal oscillations of the clump in a bucket which is due to its synchrotron motion. For a case like this one can estimate $f_{sy}$ quite easily. However, if the clump is very close to the bucket *separatrix* then one cannot use the small angle approximation, Eq. (3) for voltage measurements. For bunches with nearly uniform distribution of beam particles or multiple clumps randomly distributed, the synchrotron oscillation spectrum will be very broad and cannot be used for RF voltage calibration studies. In any case, one needs to perform FFT for each bunch separately.

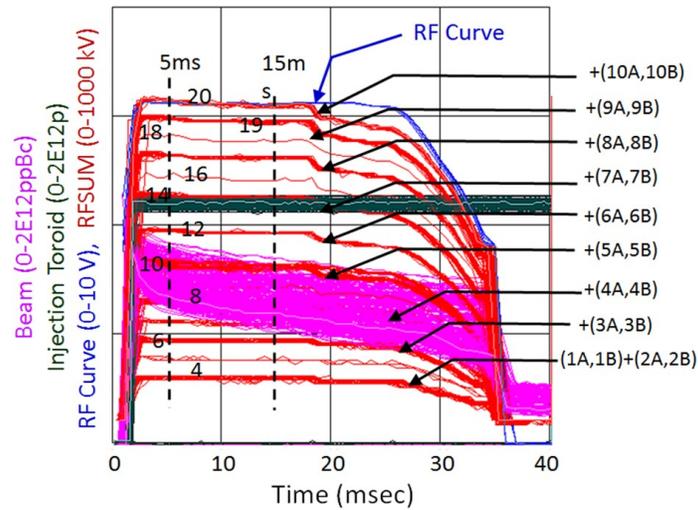

Figure 5: Beam (purple), input curve (blue), *RFSUM* (red) as we turn on more cavities for synchrotron frequency measurements while Booster running in DC mode at injection energy. The ranges of vertical scale for each parameter are also indicated. (Notice that there is about 50 kV offset to the *RFSUM* shown in this data).

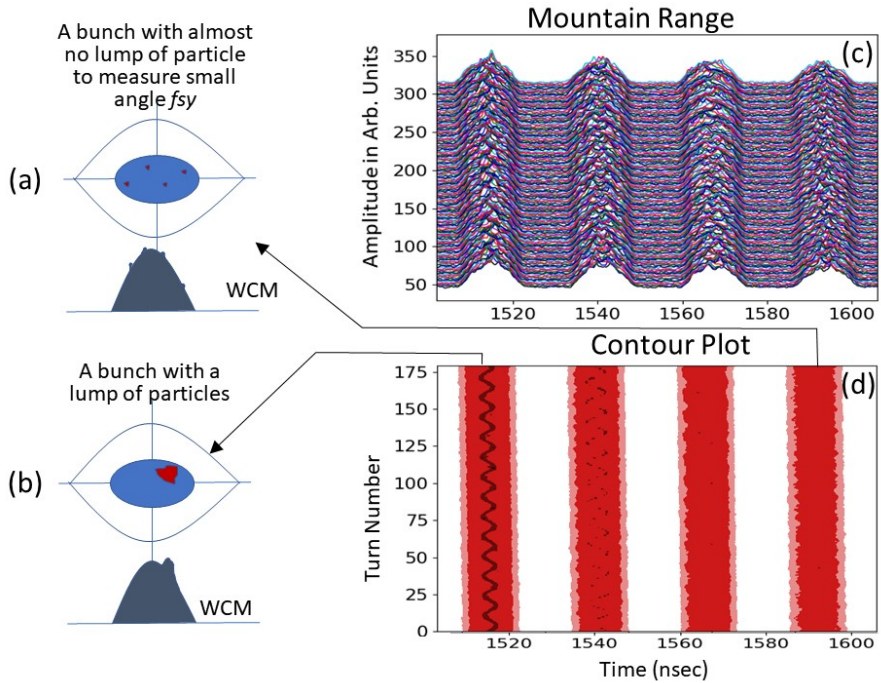

Figure 6: (a) A typical bunch with nearly uniform distribution of beam particles and (b) that with a clump of particles with their line-charge distributions, (c) Typical mountain range from WCM data and (d) the corresponding contour plots for four bunches in the Booster after beam capture. There is one turn delay between consecutive traces in the mountain range data. The bunch with a clump of particles distinctly displays a sinusoidal wave in contour plot.

Accelerators/ Beam Based RF Voltage Measurements & Longitudinal Beam Tomography …

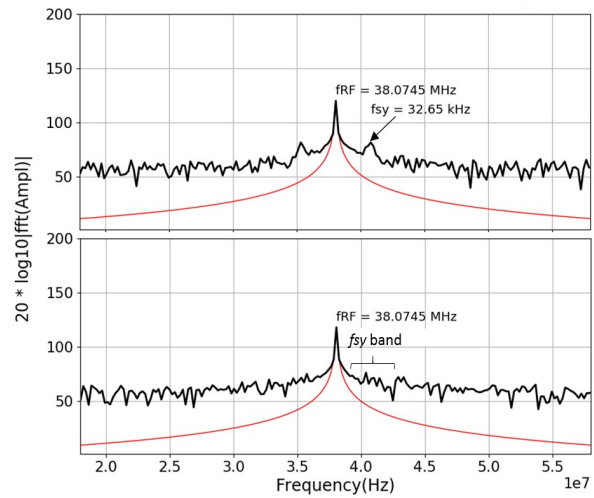

Figure 7: Fourier spectrum of a bunch at 400 MeV with a distinct clump of particles performing synchrotron oscillations (top) and that of a bunch with no clump of particles (bottom). The spectrum on the top shows distinct synchrotron side band symmetric to the bunch RF frequency of ~38 MHz, in the latter case the synchrotron side band is smeared.

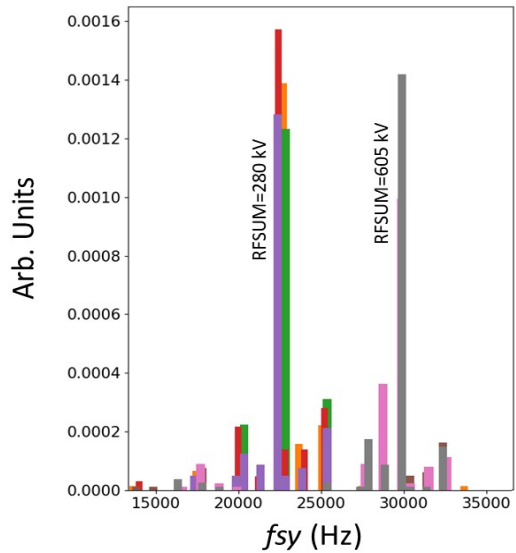

Figure 8: Sample histograms of measured $f_{sy}$ for four data sets each at two values of *RFSUM*.

The raw WCM data is rearranged to extract bunch-by-bunch synchrotron oscillation frequency. Figure 7 shows typical FFT for two bunches one with a clump of particles executing small angle synchrotron oscillations and another with almost no clumps; the indicated band of $f_{sy}$ in Fig. 7 (bottom) perhaps covers the entire region of the synchrotron frequency spectrum (*e.g.,* see Fig. 4(b)) for the 400 MeV case. Such cases and cases with large angle synchrotron frequency oscillations need to be excluded in our analysis. Sample histograms of measured $f_{sy}$ are shown in Fig. 8 for two values of *RFSUM*. The data clearly show prominent clustering for each case. By removing the outliers the mean $f_{sy}$ and its RMSW are determined. The measured $f_{sy}$ as a function of *RFSUM* at 400 MeV are presented in Fig. 9. A linear fit to the data shows that

$$V_{rf} = 0.81(\pm 0.08) \times RFSUM \qquad (4).$$

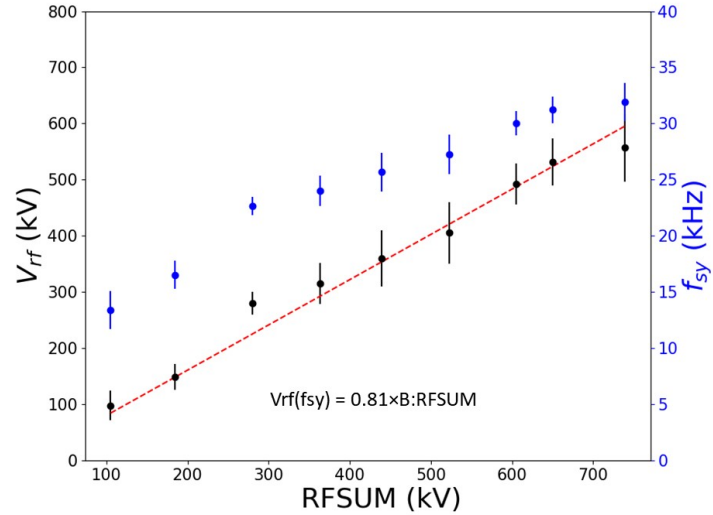

Figure 9: Measured $f_{sy}$ and $V_{rf}$ from the 400 MeV data. Only statistical errors are shown.

### 3.2. Measurements at Extraction

The method adopted to measure $f_{sy}$ at extraction is quite different from the one used at 400 MeV. In this case, the Booster *B*-field was changing all the time. Furthermore, the bunches at extraction are quite short and any clumps of particles produced at capture will have lost their identity during transition crossing. Here we should find a way to create a stationary bucket and one must look for small angle dipole oscillations of the bunches or induce such bunch oscillations in a controlled way. We tried both ways during our experiment.

About 3 msec before the end of the beam cycle of the normal operation of the Booster, the RF frequency at beam extraction is locked to that of the downstream accelerator and the radial feed-back system used for beam acceleration is turned off. Furthermore, a millisecond after the transition crossing quad dampers and mode-1 dampers are turned on to make the beam longitudinally more stable. Eventually, a bunch rotation is performed to reduce the momentum spread of the beam and match them with the RF buckets of the down-stream machines. We took advantage of many of these features of operation for our measurements. A simple combination of holding the RF frequency and turning off the radial feed-back system made the beam to be held at a constant energy. These two conditions force the beam to be in a pseudo flat-top. Holding the RF voltage to a constant value and turning off bunch rotation near extraction creates a longitudinally more stable beam. During this time, the beam moves radially inside until the *B*-field reaches its maximum value and then to outside from the central orbit as *B*-field decreases. A small angle synchrotron oscillation of the bunch can be induced by two different methods: i) give a longitudinal kick to the beam bunches using a Vernier cavity [8] or ii) turn off the mode one damper and let the dipole oscillation induced by the transition crossing-phase-mismatch survive. During our experiments, we noticed that the first method produced oscillations somewhat randomly; not every kick could produce noticeable oscillations. However, the 2nd method produced the right amount of kicks if the bunch intensity was not big enough to introduce other instabilities. Using an optimal amount of beam was also essential to avoid any effects on RF systems arising from beam loading. We used about 2E12 ppBc for these studies.

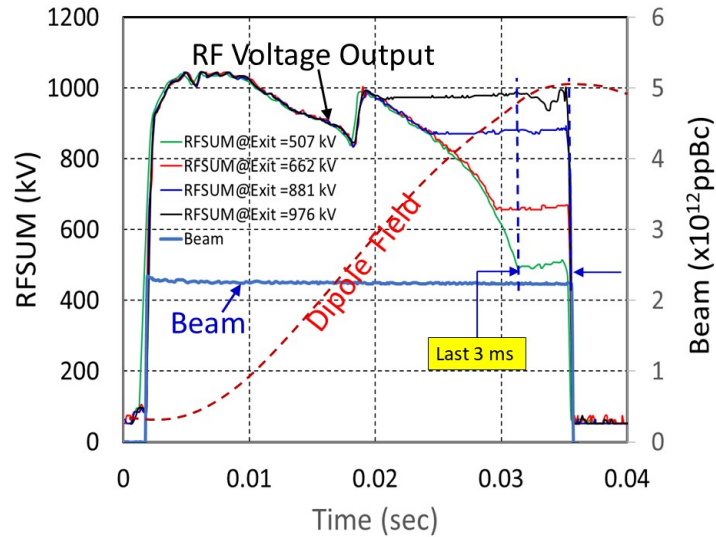

Figure 9: *RFSUM* and Beam intensity versus time of the beam cycle. During the last 3 msec before the beam extraction the *RFSUM* is fixed to different values as indicated. The expected Booster dipole ramp (dashed red curve, in arbitrary units) is also shown for clarity.

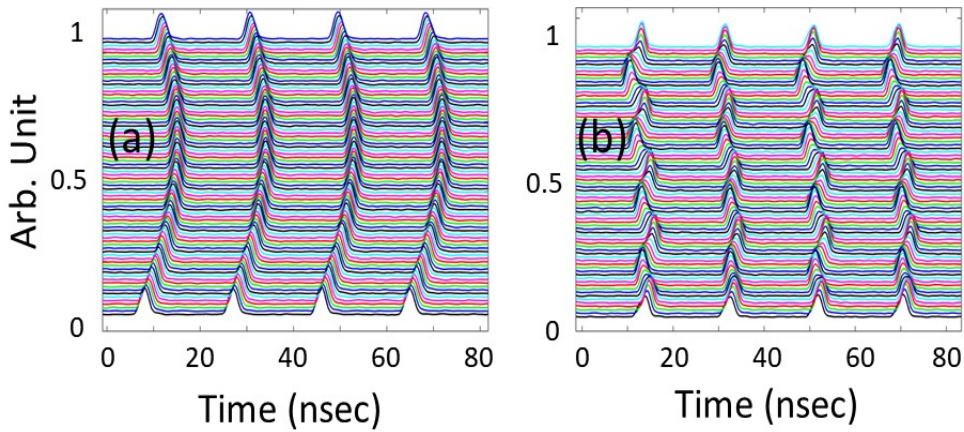

Figure 10: Comparison between mountain range data for four low beam intensity bunches with (a) mode-1 longitudinal dampers turned on and (b) damper off. For detail see the text.

Accelerators/ Beam Based RF Voltage Measurements & Longitudinal Beam Tomography …

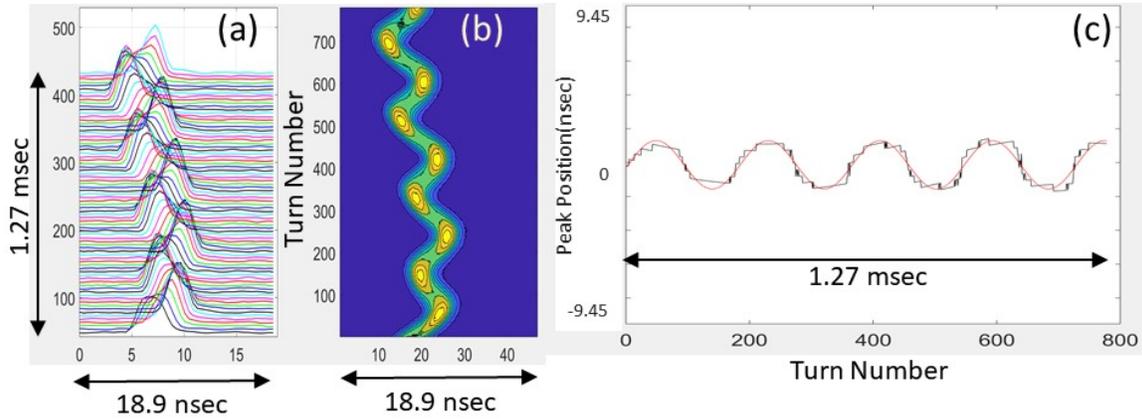

Figure 11: (a) Typical mountain range for a single bunch at extraction, (b) contour plot for the same bunch. The horizontal axis represents total span of a RF bucket of 18.9 ns. (c) peak position of the bunch versus turn number, as it oscillates in a bucket.

Figure 9 displays measured *RFSUM*, beam intensity in the Booster ring and a schematic of the dipole *B*-field from 400 MeV to 8 GeV. A kink in the *RFSUM* at around 18 msec in the cycle is introduced to minimize the momentum spread of the beam at transition energy. The efficiency of beam transmission is almost 99% through the cycle as shown in the figure. The *RFSUM* was set to the same value for all measurements until about 20 msec in the cycle and was held at a different fixed value of interest for synchrotron frequency measurements. Here the *RFSUM* is varied by changing voltage input to each of the twenty RF stations, but, not by changing the phase angle between voltage vectors or by turning off RF stations (as in the case of 400 MeV measurements). Analytical calculations using Eq. (1) suggested that the $f_{sy}$ is in the range of 1-4 kHz at extraction energy if *RFSUM* is in the range of 500 kV to 980 kV. Therefore, we expected about 3 to 12 synchrotron oscillations during the last 3 msec in the cycle, which is sufficient to determine $f_{sy}$ with good accuracy. The WCM data for the beam bunches and *RFSUM* data are collected about 2 msec prior to the beam extraction. Figure 10 compares the mountain range data of the beam for two cases. In Fig. 10(a), the mode-1 Booster longitudinal dampers were turned on and the beam was send to the RR beam dump after locking the RF frequency to that of RR. On the other hand, in Fig. 10(b), the dampers were turned off and beam was sent to the Booster beam dump. The slow motion of the bunch centroid initially to the right and later to the left in Fig. 10(a) is partly due to the changing magnetic field and beam cogging to transfer the beam to the RR RF buckets. The cogging was turned off while data in Fig. 10(b) was taken. The amplitude of observed dipole oscillations of the bunch centroids shown in Fig 10(b) were found to be about $25^0$, far better than $60^0$. To extract the bunch oscillation frequency, we fit the position of the peak intensity of a bunch in a bucket as a function of turn number according to,

$$F(t) = a_1 \sin(2\pi a_2 t + a_3) + a_4 + a_5 \sin(2\pi a_6 t + a_7) + a_8 \qquad (5).$$

where, $a_1$ and $a_5$ are amplitudes of oscillation, $a_3$ and $a_7$ are phase offsets, $a_4$ and $a_8$ are amplitude offsets and $a_2$ and $a_6$ are frequencies. The quantity $t$ is turn number or *real time*. The parameter $a_6$ is the measure of $f_{sy}$. Figure 11 illustrates a typical case of WCM data and fit to the peak position of a bunch executing small angle oscillations. A MATLAB program was developed for the complete analysis of the WCM data to extract $f_{sy}$. Figure 12 shows the measured $f_{sy}$ and $Vrf$ at extraction as a function of *RFSUM*. The statistical errors in the measured $f_{sy}$ are also presented here. A linear fit to $Vrf$ as a function of *RFSUM* with intercept enforced to $Vrf = 0$, gives

$$Vrf = 0.93\,(\pm 0.01) \times RFSUM \tag{6a}$$

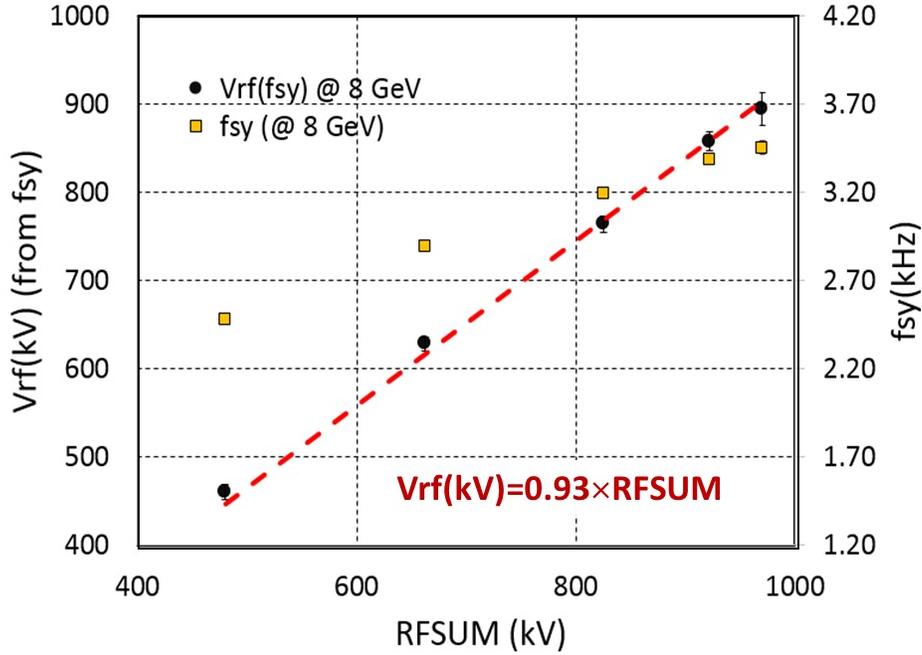

Figure 12: Measured $f_{sy}$ and $Vrf$ from the data at 8 GeV.

On the other hand, a simple linear fit give,

$$Vrf = 0.89(\pm 0.01) \times RFSUM + 37.45(\pm 0.50) \tag{6b}$$

### 3.3. Results and Discussions

There is noticeable difference between $Vrf$ measurements at 400 MeV and 8 GeV. The 400 MeV results exhibit about 10% statistical error in the measured $f_{sy}$, whereas at 8 GeV the error is <1.5%. Even though we discard outliers in the 400 MeV data, it needs additional attention with selection of the beam bunches to ensure small angle synchrotron oscillation. Such selection process is a challenging task. Furthermore, there seems to be a large systematic error in this case; we estimate that it is about 6.5% in $f_{sy}$ for synchrotron oscillation amplitude of $60^0$. On the other hand, we found out that the amplitude of oscillations in the case of 8 GeV data is about $25^0$, which gives rise to a systematic error of ~1% as shown in Fig. 13. This trasnalates to about 2% error in the measured $Vrf$. In any case, the analysis at 400 MeV and 8 GeV show that the *RFSUM* is over estimated by more than 5% relative to the true value of RF voltage.

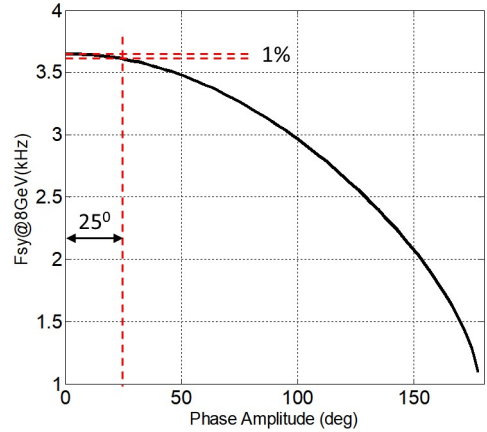

Figure 13: $f_{sy}$ versus amplitude of phase oscillation calculated for 8 GeV case on the Booster. Observational error that one could make in the measurements of $f_{sy}$ is shown.

## 4. BEAM TOMOGRAPHY

Longitudinal beam tomography [5, 6] techniques to extract (*dE*, *dt*)-phase space distribution in synchrotrons are well known and are used at various accelerator laboratories around the world. They provide important beam diagnostics. To perform longitudinal tomography one needs multiple time projection pictures of the line-charge distribution for at least half the synchrotron oscillation period, the beam phase relative to the RF wave form and $Vrf$. In the past, many attempts have been made in the Fermilab Booster [7] with little success, because the phase was changing all the time and $Vrf$ was not known to an accuracy that is needed for extracting reliable beam tomography.

As the analyses of WCM data progressed to extract the Booster RF voltage, we realized that the beam longitudinal tomography could be a spinoff of this study. The beam was always in stationary buckets with the beam phase relative to the RF wave form either at $0^0$ (for 400 MeV cases) or at $180^0$ (for 8 GeV cases), hence, the phase is a well-known quantity. Figure 13 shows the results of beam tomography carried out using software developed at CERN [6], for three different scenarios. In all these cases, we used one synchrotron oscillation length of WCM data. For example, bunch tomography depicted in Fig. 13(b) uses WCM data at 400 MeV shown in Fig. 13(a). One can see a population of non-uniform distribution of beam particles indicating non-adiabatic beam capture in the 38 MHz RF bucket. Figures. 13(d) and (e) are reconstructed (*dE*, *dt*)-phase space distributions for beam at 8 GeV, for an unstable and a stable bunch, respectively. Also notice that the bunch intensities were quite different.

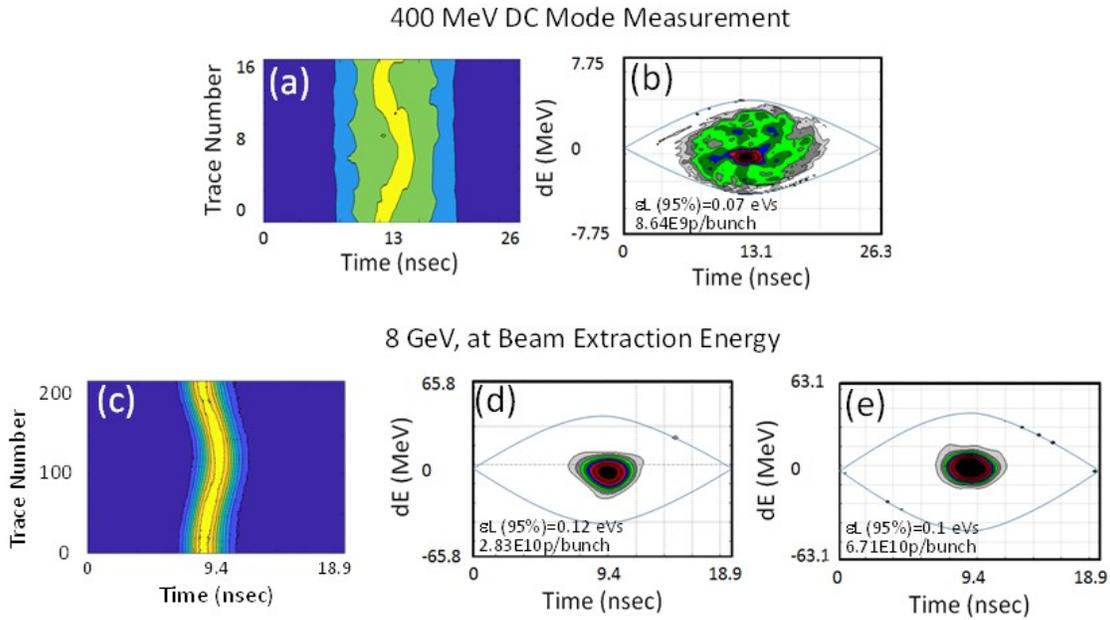

Figure 13: Displayed are, (a) contour plot of line-charge distribution for one synchrotron oscillation period from 400 MeV DC mode measurements, (b) longitudinal tomography for the data displayed in "a", (c) a sample of contour plot at extraction, (d) longitudinal tomography for low intensity beam bunch at 8 GeV with dampers off and (e) tomography for high intensity stable bunch at 8 GeV with dampers on. Bucket separatrices are also shown in "b", "d" and "e".

## 5. SUMMARY AND PROSPECTS

To meet the demands of the intensity frontier experiments at Fermilab, the accelerators including the Booster are undergoing continuous improvements. We have carried out a beam based RF voltage calibration/measurement to examine if we have enough RF power for the proposed upgrades, by measuring the small angle synchrotron frequency of the beam at 400 MeV in DC mode operation of the Booster and on a pseudo flat-top at 8 GeV. Both measurements clearly show that the measured RF voltage is at least 5% smaller than *RFSUM*, the values broadcast by Fermilab Accelerator Network System (ACNET). This deficit in the required RF voltage needs to be considered in the future intensity upgrades.

Using the WCM data, the measured RF voltage and known RF phase angle relative to the beam, we have reconstructed the longitudinal bunch tomography at 400 MeV and at an extraction energy of 8 GeV. This sort of beam tomography is first of its kind for the Fermilab Booster. We plan to develop an ACNET computer program to make beam tomography at injection and at extraction on high intensity beam cycle, which can be used as a very useful diagnostic tool to further improve the performance of the Booster.

## Acknowledgments

Authors would like to thank W. Pellico, T. Sullivan, K. Triplett and S. Chaurize, for their help at various stages of these measurements. Special thanks are due to N. Eddy from one of the authors, CMB, for his help in setting up longitudinal dampers to excite the longitudinal oscillations in the bunch. We also thank Pushpa Bhat for useful discussions. This work is supported by Fermi Research Alliance, LLC under Contract No. De-AC02-07CH11359 with the United States Department of Energy